\RequirePackage{xr-hyper}  
\documentclass[Crown,sagev,times,doublespace]{sagej}

\usepackage{changes}
\setaddedmarkup{{\color{black}#1}}
\setdeletedmarkup{}
\newcommand{\add}[1]{{#1}}

\usepackage{moreverb,url}

\usepackage[colorlinks,bookmarksopen,bookmarksnumbered,citecolor=red,urlcolor=red]{hyperref}

\newcommand\BibTeX{{\rmfamily B\kern-.05em \textsc{i\kern-.025em b}\kern-.08em
T\kern-.1667em\lower.7ex\hbox{E}\kern-.125emX}}

\usepackage{bm,dsfont}
\newcommand{\Nc}{\mathcal{N}}

\usepackage{pdfpages}

\begin{document}

\runninghead{van den Boom et al.}

\title{Bayesian inference on the number of recurrent events: A joint model of recurrence and survival}

\author{Willem van den Boom\affilnum{1,2}, Maria De Iorio\affilnum{2,3,4} and Marta Tallarita\affilnum{4}}

\affiliation{\affilnum{1}Yale-NUS College, National University of Singapore, Singapore\\
\affilnum{2}Singapore Institute for Clinical Sciences, Agency for Science, Technology and Research, Singapore\\
\affilnum{3}Yong Loo Lin School of Medicine, National University of Singapore, Singapore\\
\affilnum{4}Department of Statistical Science,
University College London, UK
}

\corrauth{
Willem van den Boom, Yale-NUS College, National University of Singapore, Singapore 138527, Singapore.
}

\email{willem@yale-nus.edu.sg}

\begin{abstract} 
The number of recurrent events before a terminating event is often of interest.
For instance, death terminates
an individual's process of rehospitalizations
and the number of rehospitalizations is an important indicator
of economic cost.
We propose a model in which the number of
recurrences before termination is a random variable of interest,
enabling inference and prediction on it.
Then, conditionally on this number, we specify a joint distribution for recurrence and survival.
This novel conditional approach induces dependence between recurrence and survival,
which is often present, for instance due to frailty that affects both.
Additional dependence between recurrence and survival is introduced by the specification of a joint distribution on their respective frailty terms.
Moreover, through the introduction of an autoregressive model, our approach is able to capture the temporal dependence in the recurrent events trajectory.
A non-parametric random effects distribution for the frailty terms
accommodates population heterogeneity and allows for data-driven clustering of the subjects.
A tailored Gibbs sampler involving reversible jump and slice sampling steps implements posterior inference.
We illustrate our model on \added{colorectal cancer} data, compare its performance with existing approaches and provide appropriate inference on the number of recurrent events.
\end{abstract}


\maketitle

\section{Introduction}
\label{s:intro}

Recurrent events arise in many applications including, amongst others, medicine, science and technology.
Typical 
examples are given by recurrent infections, asthma attacks, hospitalizations, product repairs, and machine failures.
Often, the number of recurrent events before termination of the process, such as by death or failure, is of interest.
For instance, rehospitalizations are a major financial burden for the health system\cite{Jencks2009}
and their number is used in policy making.\cite{McIlvennan2015,Zuckerman2016}
This work proposes a model which explicitly accounts for the number of recurrent events before termination
by building on recent advances in joint modelling of recurrence and termination.

%
The two main statistical approaches to inference on recurrent events are (1) modelling the intensity or hazard function of the event counts process and (2) modelling the sequence of times between recurrent events, known as gap times or waiting times.\citep{Cook2007}
The first approach is most suitable when individuals frequently experience the recurrent event of interest and the occurrence does not alter the process itself.
Here, we mention some examples that consider the dependence between recurrence and survival time.
Liu et al.,\cite{Liu2004} Rondeau et al.,\cite{Rondeau2007} Ye et al.,\cite{Ye2007} Huang et al.,\cite{Huang2010} Sinha et al.\cite{Sinha2008} and Ouyang et al.\cite{Ouyang2013}
model the intensity of the recurrent events and the survival time. The latter two approaches propose Bayesian methods with an emphasis on modelling the risk of death and the risk of rejections for heart transplantation patients.
Olesen and Parner,\cite{Olesen2006} Huang and Liu,\cite{Huang2007} Yu and Liu,\cite{Yu2011} Bao et al.\cite{Bao2012} and Liu et al.\cite{Liu2015}
model the hazard function of the recurrent events and of the survival jointly, with the recurrent events and the survival being independent conditionally on frailty parameters.
Yu et al.\cite{Yu2013} model the intensity of the recurrent events and the hazard function of survival jointly while assuming independent censoring before death.

The second approach, which focuses on the sequence of gap times, is more appropriate when the recurrent events are relatively infrequent,
when individual renewal takes place after an event,
or when the goal is prediction of the time to the next event.
Therefore, this approach is highly relevant for biomedical applications.
For instance, major recurrent cardiac events for one patient are often rather infrequent from a statistical viewpoint.
Also, healthcare planning can benefit from time-to-event predictions, especially if events require hospitalization.
Nonetheless, there is less existing work on the second approach than on the first.
This work places itself within the second framework, as the events in our application are infrequent but measured on many individuals.

Using the second approach,
Li et al.\cite{Li2018,Li2019} define survival functions via a copula to allow for dependence between recurrence and survival conditionally on frailty parameters.
Paulon et al.\cite{Paulon2018} and Tallarita et al.\cite{Tallarita2020}
propose Bayesian non-parametric models for the gap times, with the former also considering dependence with survival time.
We take this previous work as a starting point with some important differences.
Our autoregressive process for gap times has a constant mean, unlike Tallarita et al.,\cite{Tallarita2020}
as the process includes regression coefficients which would otherwise be hard to interpret.
Unlike Paulon et al.,\cite{Paulon2018} we do not assume independence of gap times conditionally on random effects, and we have separate random effects for gap and survival times, enabling greater flexibility in capturing dependence or lack thereof between recurrence and survival.

The main novel methodological contribution of this work is to explicitly enable inference on the number of recurrent events preceding a termination event.
Previous work \citep{Olesen2006,Huang2007,Li2018,Li2019,Paulon2018}
also
considers termination of the observed recurrence process.
In that work, a large number of recurrent events are 
assumed to exist for each individual with the recurrent event process defined also after the terminal event has occurred.
Then, the gap times are censored either by the survival censoring time or by the actual survival time.
The contribution to the likelihood of the gap times after censoring is set equal to one
such that
the assumed large number of censored gap times does not affect inference.
It is preferable to avoid the
often unrealistic assumption of a large  arbitrary number of recurrent events
or the continuation of the recurrence process beyond the terminal event.
More importantly, these approaches prevent inference on the number of events before termination.
This can constitute a major limitation in many applications.
For example, there is a large literature\cite{Kansagara2011,Futoma2015,Mahmoudi2020}
for hospitalizations on assessing the risk of recurrence for a time window such as within 30 days of hospital discharge
as this has serious implications for healthcare cost.
Our model can not only provide such risk estimates but is also able to infer the number of rehospitalizations which can aid healthcare planning (e.g.\ budgeting, provisioning).

We overcome the limitations of previous approaches by adopting a conditional approach.
More in detail, 
we explicitly model the number of recurrent events before the terminal event.
Then, conditionally on the number of events, we specify a joint distribution for gap times and survival. To the best of our knowledge, this modelling strategy has never been employed in the context of joint modelling of survival and recurrence processes.
The resulting posterior inference is computationally more challenging than the models proposed in the available literature. We therefore develop a Markov chain Monte Carlo algorithm which relies on various computational tools such as reversible jump Markov chain Monte Carlo \citep{Green1995} and slice sampling.\citep{Neal2003}
Our explicit modelling of the number of recurrent events
yields a more intuitive model specification,
and captures the dependence structure between the recurrence and survival processes, which is important in medical applications and beyond.
The main motivating result though is intuitive inference and prediction for the number of recurrent events.

An important factor in medical applications related to recurrent events and survival time is the overall frailty.
Increased frailty is often associated with both increased disease recurrence and reduced survival.
Subject-specific random effects describe the frailty
by informing both the survival time and the dependence of subsequent gap times.
The random effects are modelled flexibly with a Dirichlet process \citep[DP,][]{Ferguson1973} prior
as in Paulon et al.\cite{Paulon2018} and Tallarita et al.\cite{Tallarita2020}
It is well known that the DP is almost surely discrete.
This feature is particularly useful in applications as it allows for data-driven clustering of observations. 
If $G$ is $\text{DP}(M,\, G_0)$ with concentration parameter $M$ and base measure $G_0$, then it admits a stick-breaking representation \citep{Sethuraman1994} and can be represented as
\[
	G(\cdot) = \sum_{h=1}^\infty w_h\, \delta_{\bm\theta_h}(\cdot)
\]
where $\delta_{\bm\theta_h}$ is a point mass at $\bm\theta_h$, the weights $w_h$ follow the stick-breaking process $w_h = {V_h \, \prod_{j<h}(1-V_j)}$ with $V_h\overset{\text{i.i.d.}}{\sim}\text{Beta}(1,\, M)$,
and the atoms $\{\bm{\theta}_h\}_{h=1}^\infty$ are such that $\bm\theta_h \sim G_0$. The sequences $\{\bm{\theta}_h\}_{h=1}^\infty$ and $\{ V_h \}_{h=1}^\infty$ are independent. 
The discreteness of $G$ induces clustering of the subjects, based on the unique values of the random locations $\bm\theta_h$, where the number of clusters is learned from the data. This choice allows for extra flexibility, variability between individual trajectories, overdispersion and clustering of observations, and overcomes the often too restrictive assumptions underlying a parametric random effects distribution.

Paulon et al.\cite{Paulon2018} specify a single random effects parameter which influences both the distribution of the gap times and the distribution of survival.
Instead, we introduce different random effects parameters, one for the recurrence process and one for the survival.
We model these jointly using a DP prior,
ensuring dependence between recurrence and survival.
Additionally, we specify 
an autoregressive model for the gap times to capture the dependence between subsequent gap times as some persistence of recurrence across time is expected.
Tallarita et al.\cite{Tallarita2020} also use an autoregressive model, but on the random effects instead of the gap times themselves.

The paper is structured as follows. Section~\ref{s:model} introduces the model.
Section~\ref{sec:simulation} checks its performance on simulated data.
Section~\ref{sec:readmission} discusses an application to \added{colorectal cancer} data. Section~\ref{s:comparison} compares our approach with existing ones. Section~\ref{s:discuss} concludes the paper.

\section{Model}
\label{s:model}

\subsection{Notation}
\label{sec:notation}

We consider data on $L$ individuals. Let
$T_{i0}$ denote the start time of the recurrent event process for individual $i$.
We assume $T_{i0} = 0$ for $i=1,\dots,L$.
Let $S_i$ denote the survival time for individual $i$ since the start of the corresponding event process.
Each individual $i$ experiences $N_i$ recurrent events over the time interval $(0,\, S_i]$.
Let $T_{ij}$ denote the $j$th event time for individual $i$. Then, the last event time $T_{i N_i}$ is less than or equal to $S_i$.

Some event processes are right censored due to end of study, as in the application of Section~\ref{sec:readmission}, or loss to follow-up.
We assume completely independent censoring.
This contrasts with the survival time $S_i$ which our model allows to depend on the event process.
Let $c_i$ denote the minimum of the censoring time and the survival time $S_i$ for individual $i$, who is thus observed over the interval $(0,\, c_i]$.
Let $n_i\leq N_i$ denote the number of events that are observed over the interval $(0,\, c_i]$.
Either $S_i$ or the censoring time is observed.
If $S_i$ is observed, then $N_i = n_i$ and
$0 < T_{i1} < \cdots < T_{i N_i} \leq S_i = c_i$.
If $S_i$ is not observed, then $N_i\geq n_i$ and $S_i > c_i$ are unknown and object of inference. In this case,
$0 < T_{i1} < \cdots < T_{i n_i} \leq c_i < T_{i (n_i+1)} < \cdots < T_{i N_i} \leq S_i$.
We define the log gap times as
\begin{equation} \label{eq:def_Y}
	Y_{ij} = \log(T_{ij} - T_{i(j-1)}),
\end{equation}
for $j = 1,\dots,N_i$.
The $q$-dimensional vector $\bm x_i$ contains individual-specific covariates.

\subsection{Likelihood specification}
\label{sec:likelihood}

Firstly, we assume that the number of gap times $N_i$ follows a \replaced{negative binomial}{Poisson} distribution with \replaced{shape parameter $r$ and mean}{rate parameter} $\lambda$ \deleted{that is truncated by $N_i\geq 1$:
$N_i \sim \mathds{1}_{[1,\infty)}\,\text{Poisson}(\lambda)$,}
independently for $i=1,\dots,L$.
Conditionally on $N_i$, we specify a joint model for the log gap times $\bm Y_{i}=(Y_{i1},\ldots, Y_{iN_i})^T$ and the survival time $S_i$.
We define the joint density ${p(\bm Y_i, S_i\mid \text{\textemdash} )}$ 
where the joint space is constrained by   $T_{i N_{i}} \leq S_i$.
This induces dependence among $N_i$, $\bm Y_i$ and $S_i$ in a principled manner.
We build on existing literature \citep{Paulon2018,Tallarita2020} by assuming that the gap times and survival times follow truncated log-normal distributions where the pairs $(\bm Y_i, S_i)$
are mutually independent for $i = 1,\dots,L$,
conditionally on the random effects,
number of recurrent events $N_i$ and the other parameters in the model.
Conditionally on $N_i$, the joint distribution for $(\bm Y_i, S_i) $  contains two components:
\begin{equation} \label{eq:joint}
	p(\bm Y_i, S_i\mid \text{\textemdash} ) \propto f(\bm Y_i \mid \text{\textemdash} )\, f(S_i \mid \text{\textemdash} )\,  \mathds{1}(T_{i N_{i}} \leq S_i)
\end{equation}
one defining the recurrence process and one for the survival, where
$T_{iN_i} = \sum_{j = 1}^{N_i} e^{Y_{ij}}$ per \eqref{eq:def_Y}. The dependence between the two processes is defined through the constraint on the joint space  and the specification of  a joint random effect distribution on the process specific parameters.

The random effects parameter for the gap times is the two-dimensional vector $\bm m_i$ which characterizes an autoregressive model for $\bm Y_i$ that captures dependence between subsequent gap times:
\begin{multline} \label{eq:gap_times}
	f(\bm Y_i \mid \bm\beta, \bm m_i, N_i, \sigma^2, \bm x_i) =
	\mathcal{N}(
		Y_{i1} \mid
		\bm x_i^T \bm\beta + m_{i1},\, \sigma^2
	)\\ \times
	\prod_{j = 2}^{N_i}
	\mathcal{N}\{
		Y_{ij} \mid
		\bm x_i^T \bm\beta + m_{i1} + m_{i2} (Y_{i(j-1)} - \bm x_i^T \bm\beta - m_{i1}),\, \sigma^2
	\}
	,
\end{multline}
for
$N_i\geq 1$. When $N_i=0$,  then  we assume that $ f(\bm Y_i \mid \bm\beta, \bm m_i, N_i, \sigma^2, \bm x_i) $ is equal to a constant.
The $q$-dimensional vector $\bm\beta$ consists of covariate effects on the gap times.
This resembles the autoregressive model on the random effects in Equation 2 of Tallarita et al.\cite{Tallarita2020} Two main differences are 
due to the fact that
the the mean of $Y_{ij}$ is the same for all $j$ conditionally on the remaining parameters in our model
and that
Tallarita et al.\cite{Tallarita2020} do not consider a survival process, which implies
the truncation $T_{iN_i} \leq S_i$ in our work. 
The truncation results from our conditioning on $N_i$ whereas existing literature \citep{Paulon2018,Tallarita2020,Aalen1991} specifies the likelihood as a joint distribution of
the number of events $n_i$ observed over the interval $(0,\, c_i]$
and their log gap times $Y_{i1},\dots, Y_{in_i}$.
The regression coefficient $\bm\beta$ in \eqref{eq:gap_times} has the usual interpretation since the mean of $Y_{ij}$ equals $\bm x_i^T \bm\beta + m_{i1}$ for all $j$.

The  survival component of the model is proportional to a log-normal density:
\begin{equation} \label{eq:surv_times}
	\added{f(S_i\mid \bm\gamma, \delta_i,  \bm x_i \} =  \mathcal{LN}\left\{ \added{S_i \mid} \bm x_i^T\bm\gamma + \delta_i,\, \eta^2\right\}}
\end{equation}
where the $q$-dimensional vector $\bm\gamma$ consists of covariate effects on the survival time and $\delta_i$ denotes a random effects parameter.
Covariate effects can differ between gap and survival times, for instance if a therapy delays disease recurrence but does not prolong survival.
Therefore, the model on the gap times in \eqref{eq:gap_times} and on the survival times in \eqref{eq:surv_times} have distinct regression coefficients $\bm\beta$ and $\bm\gamma$, respectively.
\added{Ultimately, due to the use of a DP prior on $(\bm m_i,\delta_i)$ our sampling model is an infinite mixture with weights and location deriving from the Dirichlet Process.\cite{Lo1984,DeIorio2009} This allows us to overcome the often too restrictive assumptions imposed by a choice of a parametric model. 
}

\subsection{Prior specification}
\label{sec:prior_spec}

We specify a non-parametric prior for 
the random effects parameters $\bm m_{i}$ and $\delta_i$ in \eqref{eq:gap_times} and \eqref{eq:surv_times}. In more detail, 
$(\bm m_{i},\, \delta_i) \sim G$ independently for $i=1,\dots,L$, where $G\sim\text{DP}(M,\, G_0)$,
with $M\sim \text{Gamma}(a_M,\, b_M)$ and base measure
$G_0 = \mathcal{N}_2(\bm 0_{2\times 1},\,\sigma^2_m \bm I_2) \times \mathcal{N}(0,\,\sigma^2_\delta)$.
Finally, the prior distributions on the remaining parameters are
$\bm\beta\sim\mathcal{N}_q(\bm 0_{q\times 1},\, \sigma^2_\beta\, \bm I_q)$, 
$\bm\gamma\sim\mathcal{N}_q(\bm 0_{q\times 1},\, \sigma^2_\gamma\, \bm I_q)$,
$\sigma^2 \sim \text{Inv-Gamma}(\nu_{\sigma^2}/2,\, \nu_{\sigma^2}\, \sigma^2_0 / 2)$,
$\eta^2 \sim \text{Inv-Gamma}(\nu_{\eta^2}/2,\, \nu_{\eta^2}\, \eta^2_0 / 2)$\added{, $r \sim \text{Gamma}(a_r,\,b_r)$} and
$\lambda \sim \text{Gamma}(a_\lambda,\,b_\lambda)$.

\subsection{Posterior inference}
\label{sec:inference}

Posterior inference is performed through a Gibbs sampler algorithm.
This includes imputing $N_i$, $Y_{i(n_i+1)},\dots,Y_{iN_i}$ and $S_i$ for each censored individual $i$ by sampling them according to the model in Section~\ref{sec:likelihood}.
The Gibbs update for $N_i$ and $\bm Y_i$ is transdimensional since the number of events $N_i$ represents the dimensionality of the sequence of log gap times $\bm Y_i$. This requires devising a reversible jump sampler \citep{Green1995} for $N_i$ and $\bm Y_i$.

Most full conditional distributions are intractable due to the truncation $T_{iN_i} \leq S_i$, which for instance causes the normalization constant
\added{of ${p(\bm Y_i, S_i\mid \text{\textemdash} )}$}
to depend on parameters of interest.
We use slice sampling \citep{Neal2003} to deal with this intractability.
The normalization constant
\added{of ${p(\bm Y_i, S_i\mid \text{\textemdash} )}$}
is also intractable. We therefore approximate it using the Fenton-Wilkinson\citep{Fenton1960} method.
Algorithm~8 from Neal\cite{Neal2000} is implemented to sample the DP parameters $(\bm m_{i},\, \delta_i)$.
Section~\ref{sec:gibbs} of the supplemental material
details and derives the Markov chain Monte Carlo algorithms.

\section{Simulation study}
\label{sec:simulation}

We investigate the performance of our model and the Markov chain Monte Carlo algorithm via a simulation study.
We consider $L = 150$ individuals spread \add{across} three clusters of size 50 each \added{by assigning every first, second and third individual to Cluster 1, 2 and 3, respectively,}
and $q=2$ covariates which are drawn independently from a standard uniform distribution.
Then, we \replaced{sample}{generate} data according to the likelihood in Section~\ref{sec:likelihood} with
$\bm\beta = \bm\gamma = (-1,1)^T$,
\added{$r=1$,}
$\lambda = 7$,
$\sigma^2 = 1$,
$\eta^2 = 1$,
$\bm m^*_h = \{h,\, 0.8\,(h-2)\}^T$
and $\delta^*_h = h + 4$
where $\bm m^*_h = \bm m_i$ and $\delta^*_h = \delta_i$ if and only if individual $i$ belongs to the $h$th cluster for $h = 1,2,3$.
From these simulated data,
we construct four different scenarios,
namely where 0\%, 50\%, 80\% and 90\% of the individuals\added{, selected uniformly at random,} are censored.
The censoring times $c_i$ are sampled uniformly from the time interval $(T_{i1},\, S_i)$
between the first event recurrence and death
similarly to the simulation in Section~4 of Tallarita et al.\cite{Tallarita2020}
\added{if $N_i \geq 1$, and from $(0,\, S_i)$ otherwise.}

\begin{figure}
\centering
\includegraphics[width=0.8\textwidth]{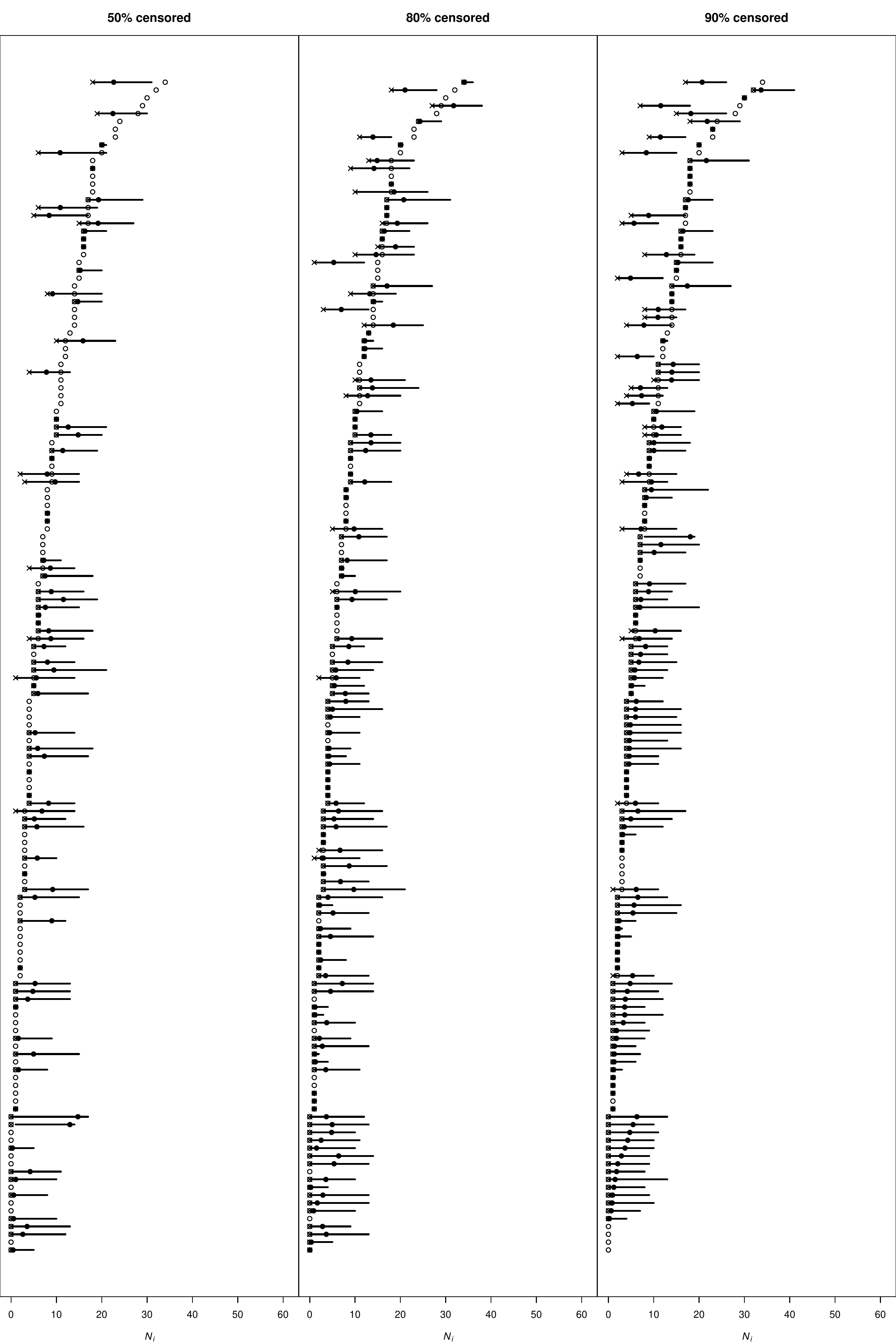}
\caption{
The number of gap times $N_i$ (circle) and, if applicable, their posterior means (dot) and 95\% posterior credible intervals (lines) for each individual from our model fitted on the simulated data.
For censored individuals, the number of observed gap times $n_i$ is marked by `$\times$'.
}
\label{fig:simul_N}
\end{figure}

\begin{figure}
\centering 
\includegraphics[width=0.8\textwidth]{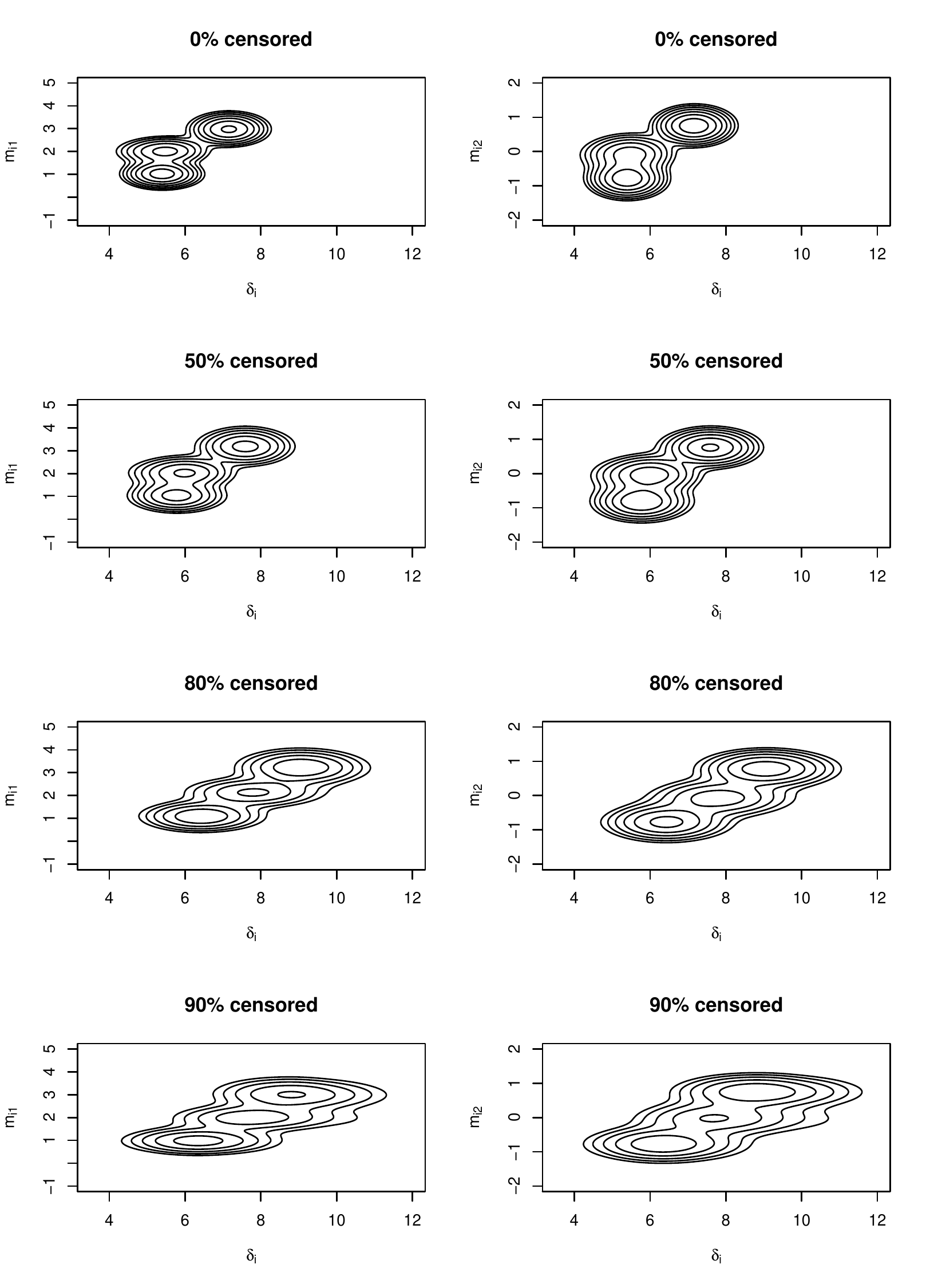}
\caption{
Contour plots of the log of the bivariate posterior predictive densities of $(m_{i1},\, \delta_i)$ (left) and $(m_{i2},\, \delta_i)$ (right) for a hypothetical new individual from the simulated data.
}
\label{fig:simul_m_new}
\end{figure}

\begin{figure}
\centering 
\includegraphics[width=0.8\textwidth]{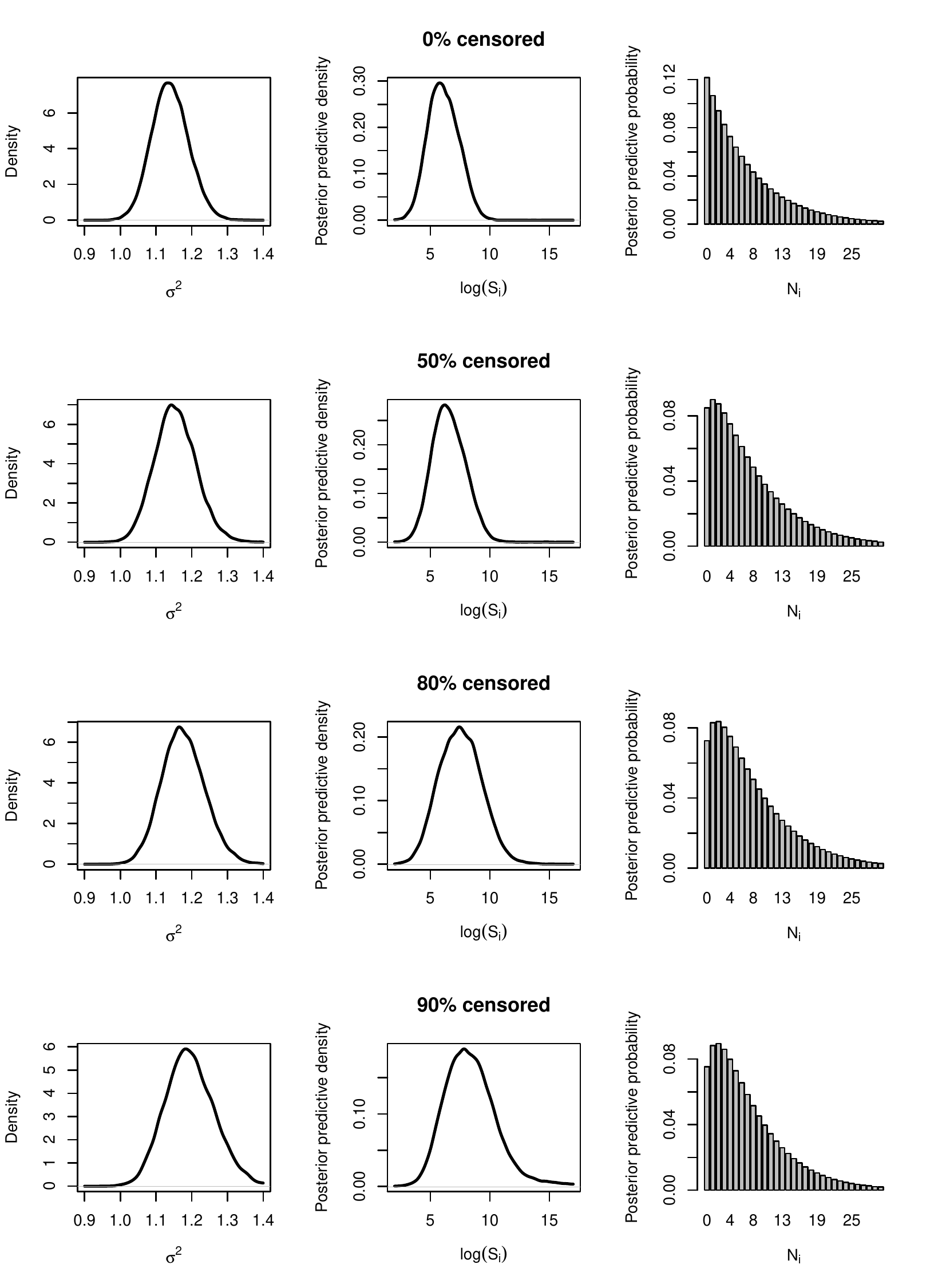}
\caption{
Posterior \replaced{density for $\sigma^2$, and posterior predictive density for $\log(S_i)$ and posterior predictive probability mass function for $N_i$ for a hypothetical new patient with covariates equal to their sample medians}{densities for $\sigma^2$, $\eta^2$ and $\lambda$} from our model fitted on the simulated data.
}
\label{fig:simul_variance}
\end{figure}

\begin{figure}
\centering
\includegraphics[width=0.8\textwidth]{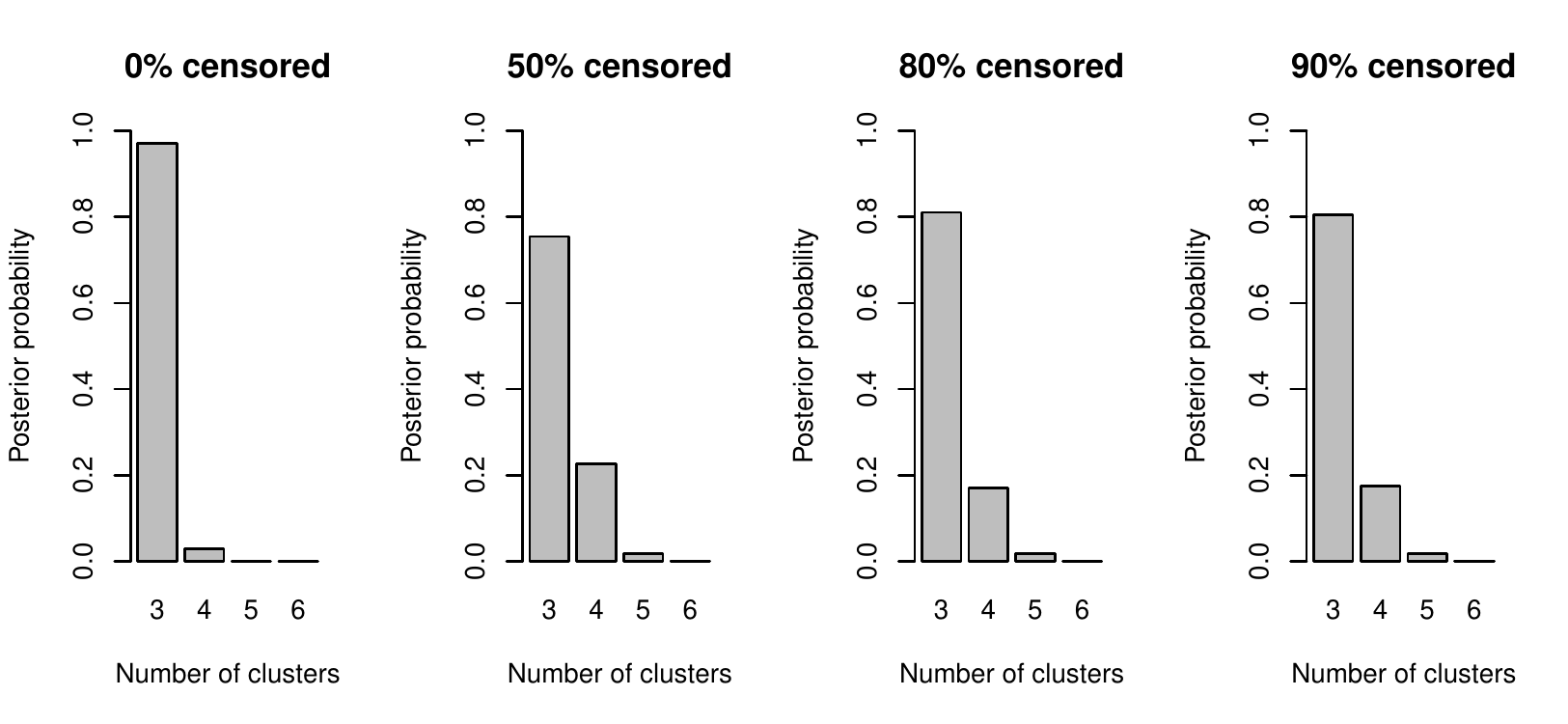}
\caption{
Posterior distribution of the number of clusters from our model fitted on the simulated data.}
\label{fig:simul_K}
\end{figure}

\begin{figure}
\centering
\includegraphics[width=0.8\textwidth]{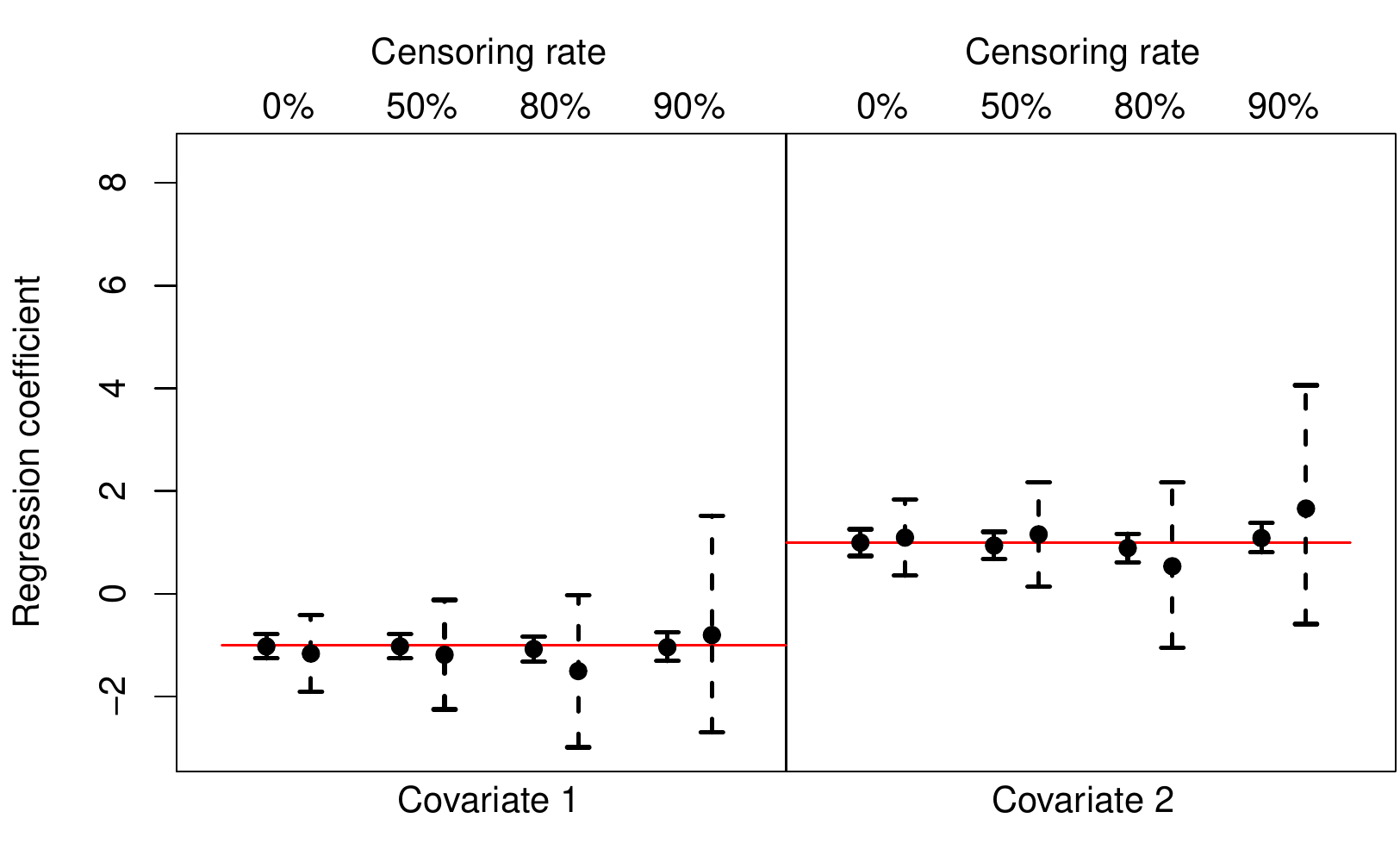}
\caption{
Posterior means (dot) and 95\% marginal posterior credible intervals (lines) of the regression coefficients from our model fitted on the simulated data. The solid lines represent credible intervals for the regression coefficients $\bm\beta$ in \eqref{eq:gap_times} for the gap times model. The dashed lines correspond with the regression coefficients $\bm\gamma$ in \eqref{eq:surv_times} for the survival times.
The horizontal line marks the true $\bm\beta = \bm\gamma = (-1,1)^T$ from which the data were simulated.
}
\label{fig:simul_coef}
\end{figure}

We choose hyperparameters yielding uninformative prior distributions as detailed in Section~\ref{sec:prior_spec_SM} of the supplemental material.
The base measure $G_0$ of the DP prior has high variance.
A priori, $\sigma^2$ and $\eta^2$ have an expected value of one and a variance of $100$.
We run the Gibbs sampler for 200 000 iterations, discarding the first 20 000 as burn-in and thinning every 10 iterations, resulting in a final posterior sample size of 18 000.

Figures~\ref{fig:simul_N} through \ref{fig:simul_coef} summarize the results.
The predictions and uncertainty quantification for the number of recurrent events $N_i$ in Figure~\ref{fig:simul_N} are sensible with only \replaced{16 (5\%)}{one} of the 330 credible intervals
not covering the true $N_i$.
Figures~\ref{fig:simul_m_new} and \ref{fig:simul_variance} show increased posterior uncertainty for higher levels of censoring.
This increase in uncertainty is larger for $\delta_i$ \deleted{and $\eta^2$}, which relate\added{s} to the survival time $S_i$,
than for other parameters which relate to the recurrence process.
A reason for this is that, for a censored individual $i$, \added{often} some event times are observed while $S_i$ is right-censored and thus not observed.
The posterior mass on the actual number of clusters is marginally higher for the uncensored than for the censored data in Figure~\ref{fig:simul_K}, although it must be noticed that posterior inference on $K$ is robust across different level of censoring.
Figure~\ref{fig:simul_coef} shows accurate inference for $\bm\beta$ with the uncertainty \added{in} $\bm\gamma$ increasing with the censoring rate
as a higher censoring induces more uncertainty about the survival time $S_i$.

\added{Section~\ref{sec:add_simul} of the supplemental material contains additional simulation studies.
Our simulations show that poster inference is robust not only to the choice of hyper-parameters of the negative binomial distribution, but also  
to model misspecification. Nevertheless, it must be noted that  posterior mean estimates and their associated  bias and  mean squared errors are affected by censoring rate.}

\section{Application to \added{colorectal cancer} data}
\label{sec:readmission}

\subsection{Data description and analysis}
\label{sec:set-up}

We apply our model to the \added{colorectal cancer data described in Gonzalez et al.\cite{Gonzalez2005}
which consider $L=403$ patients diagnosed with colorectal cancer between 1996 and 1998 in Bellvitge University Hospital in Barcelona, Spain.
The data consist of hospital readmissions related to colorectal cancer surgery up until 2002
and are available as part of the R package \texttt{frailtypack}.\citep{Rondeau2012}
The date of surgery represents the origin of a patient's recurrence process such that $T_{i0} = 0$ for all $i$.
Consequently, $n_i$ represents the number of observed gap times between subsequent hospitalizations.
Patients experience between zero and 22 hospitalizations each and $\sum_{i=1}^L n_i = 458$ in aggregate. Table~\ref{tab:ni_real}
shows how they are distributed across patients.
Gap times are defined as the difference between successive hospitalizations and, as such, capture both the length of stay in the hospital and the time between discharge and the next hospitalization.
}

\begin{table}
\caption{Frequency table of the number of observed gap times $n_i$ and, for the 109 uncensored patients, the total number of gap times $N_i$ in the colorectal cancer data. \label{tab:ni_real}}
\centering
\begin{tabular}{lcccccccccccccc}
\hline \hline
$n_i$ & 0 & $1$ & $2$ & $3$ & $4$ & $5$ & $6$ & 7 & $8$ & $9$ & $10$ & $11$ & $16$ & $22$ \\
\hline
Frequency (all) & 199 & 105 & 45 & 21 & 15 & 8 & 4 & 0 & 1 & 1 & 1 & 1 & 1 & 1 \\
Frequency (uncensored) & 36 & 33 & 16 & 10 & 6 & 4 & 0 & 0 & 1 & 1 & 0 & 0 & 1 & 1 \\
\hline 
\end{tabular}
\end{table}

The main clinical outcome of interest is deterioration to death.
We therefore define the survival time $S_i$ as the time to death.
The survival times of \added{294} out of the \added{403} recurrence processes are censored due to the follow-up ending \added{in 2022}, resulting in unobserved total number of gap times $N_i$.

\begin{figure}
\centering 
\includegraphics[width=0.8\textwidth]{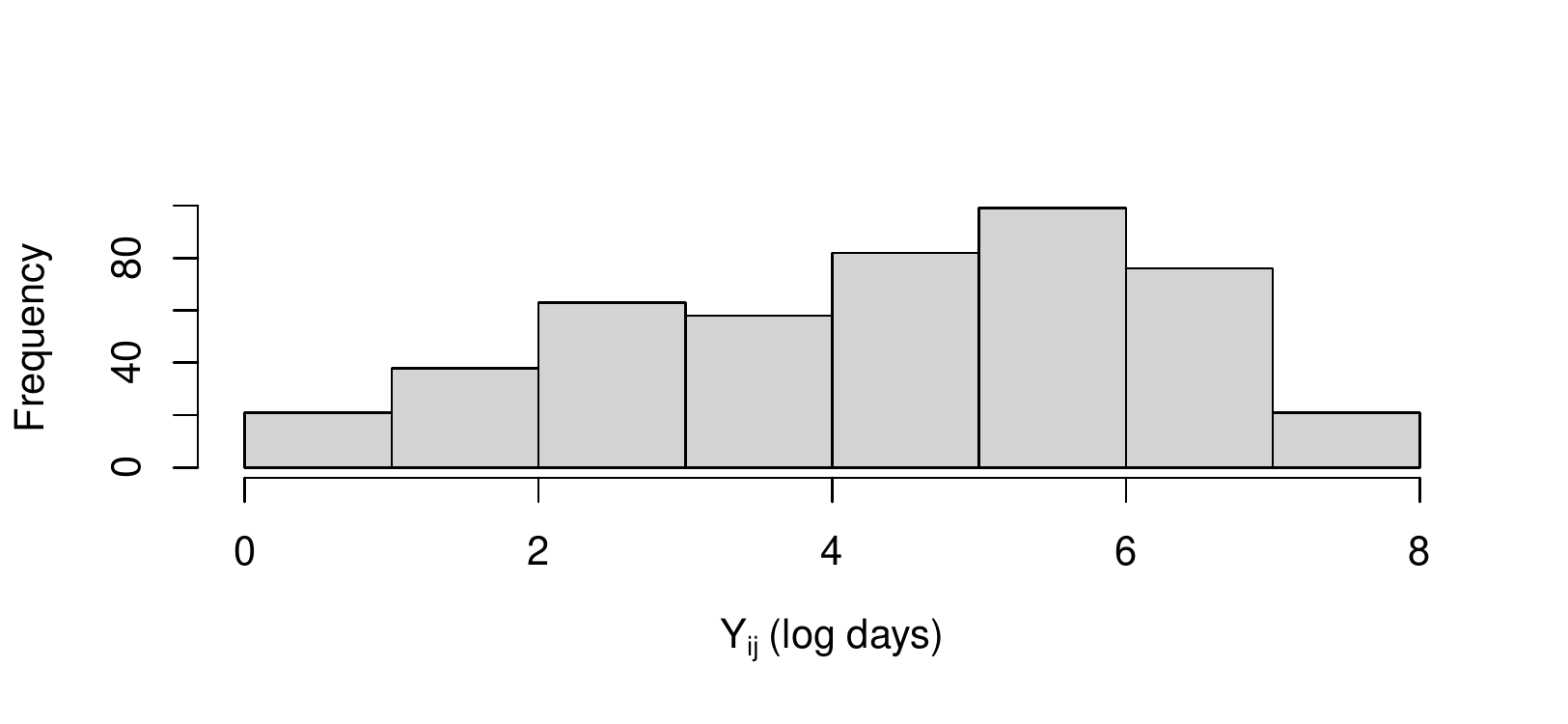}
\caption{Histogram of the 458 observed log gap times $Y_{ij}$ in the colorectal cancer data.}
\label{fig:readmission_w}
\end{figure}

\begin{figure}
\centering 
\includegraphics[width=0.8\textwidth]{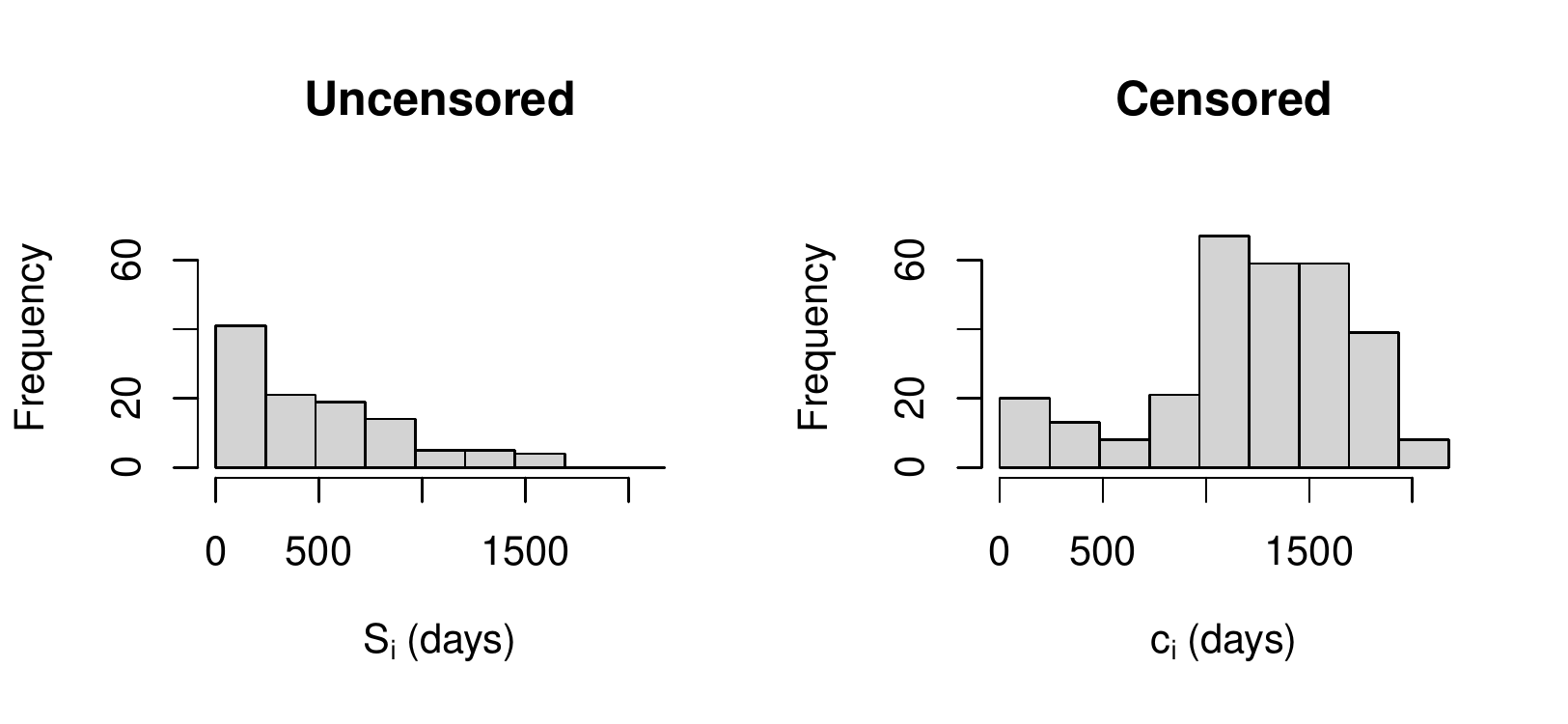}
\caption{Histograms of the log of the 109 observed survival times $S_i$ (left) and the 294 censoring times (right) in the colorectal cancer data. If the survival time $S_i$ is observed, then $c_i=S_i$.}
\label{fig:readmission_T}
\end{figure}

Patient characteristics
\added{considered are the binary variables 1) whether the patient received radiotherapy or chemotherapy and
2) gender, and Dukes' tumour classification which takes as levels stage A-B, C or D.
We dummy code Dukes' classification with stage A-B as baseline
resulting in
a subject-specific $4$-dimensional covariate vector $\bm x_i$, with $q=4$.
Table~\ref{tab:readmission_K}, and Figures~\ref{fig:readmission_w} and \ref{fig:readmission_T} summarize the patient characteristics, and the gap and survival times.}

We use the same priors, from Section~\ref{sec:prior_spec_SM} of the supplemental material, and set-up of the Gibbs sampler as the simulation study in Section~\ref{sec:simulation}.
Then, the regression coefficients $\bm\beta$ and $\bm\gamma$ have high prior variance.

\begin{figure}
\centering
\includegraphics[width=\textwidth]{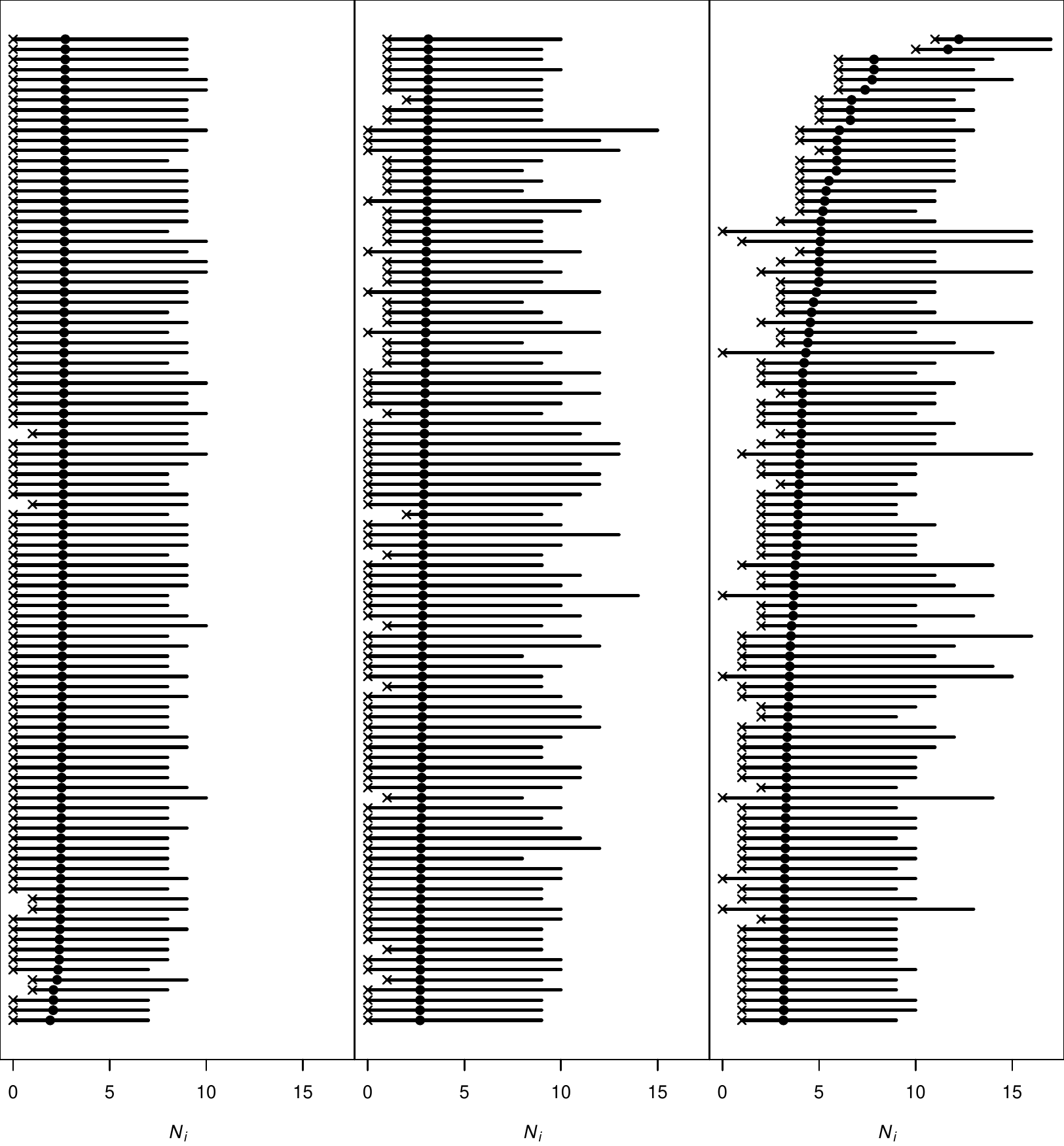}
\caption{
The posterior means (dot) of the number of gap times $N_i$ and 95\% posterior credible intervals (lines) for each censored patient from our model fitted on the colorectal cancer data.
The number of observed gap times $n_i$ is marked by `$\times$'.
}
\label{fig:readmission_N}
\end{figure}

\subsection{Posterior inference on the number of recurrent events}

Figure~\ref{fig:readmission_N} summarizes the posterior distribution of the total number of gap times $N_i$
for each \added{censored} patient.
The posterior means for the censored $N_i$ are generally in line with the observed $N_i$ \added{in Table~\ref{tab:ni_real} though the lowest posterior mean is with 1.9 higher than the lowest observed $N_i$}.
This is expected since patients with longer survival times $S_i$ are both more likely to have a higher number of gap times $N_i$ and to be censored due to end of study.
Our model flexibly captures $N_i$'s uncertainty, which varies notably across censored patients.
These findings highlight the importance of modelling and inferring $N_i$ when the number of events is censored and, therefore, unknown.

\subsection{Posterior inference on the regression coefficients}

\begin{figure}
\centering 
\includegraphics[width=0.8\textwidth]{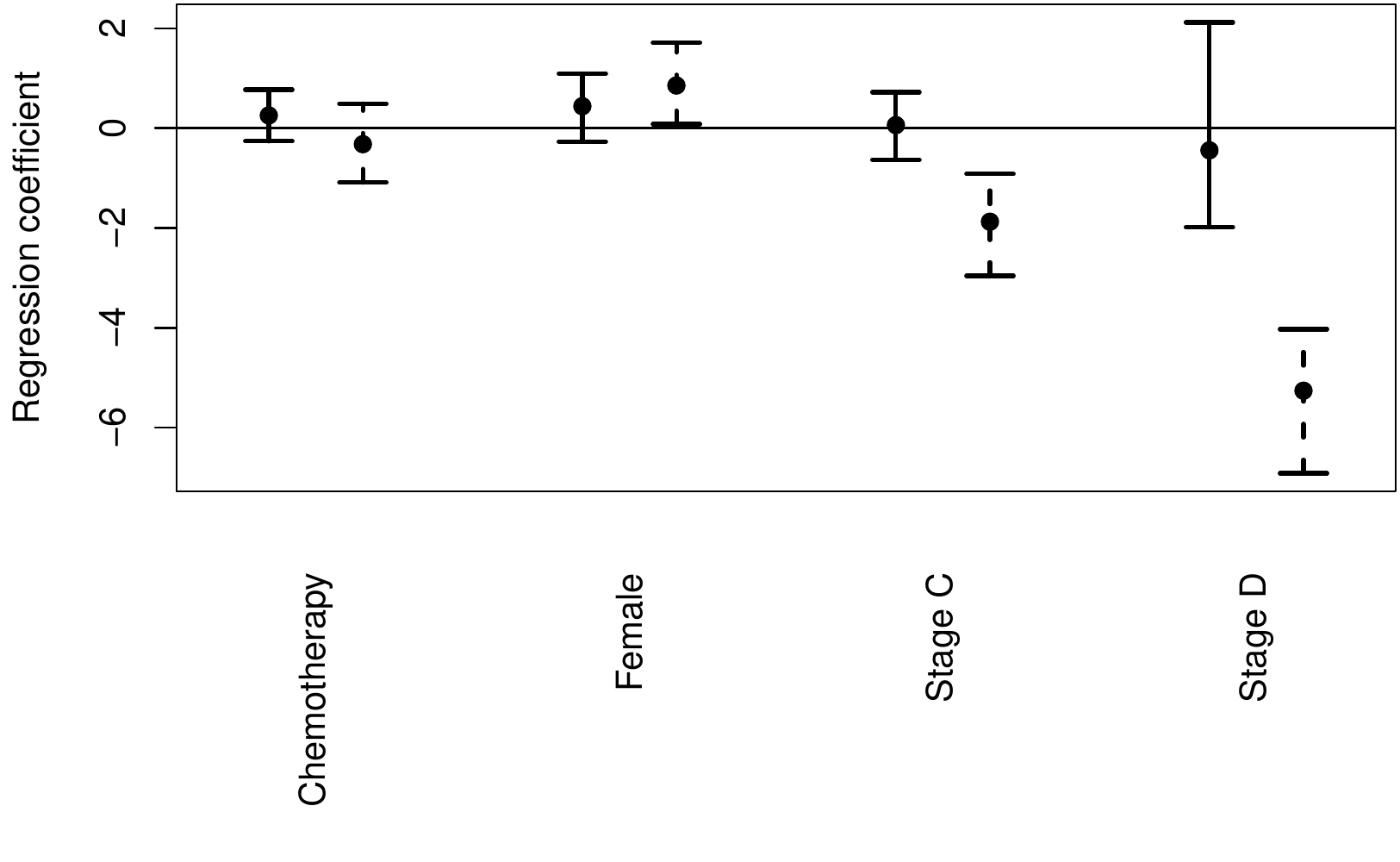}
\caption{Posterior means (dot) and 95\% marginal posterior credible intervals (lines) of the regression coefficients from our model fitted on the colorectal cancer data. The solid lines represent credible intervals for the regression coefficients $\bm\beta$ in \eqref{eq:gap_times} for the gap times model. The dashed lines correspond with the regression coefficients $\bm\gamma$ in \eqref{eq:surv_times} for the survival times.}
\label{fig:readmission_CI}
\end{figure}

The regression coefficients $\bm{\beta}$ and $\bm{\gamma}$ capture the covariate effects on the recurrent event and survival processes, respectively.
Figure~\ref{fig:readmission_CI} shows
\added{negative coefficients for the more advanced tumour stages C and D in the survival regression but not for the gap times model.
This suggests that more severe tumours, while they negatively affect survival,
do not have an effect on rehospitalization rate beyond the link between survival and hospitalization implied by our joint model.
}

\subsection{Posterior inference on the cluster allocation}

\begin{figure}
\centering 
\includegraphics[width=\textwidth]{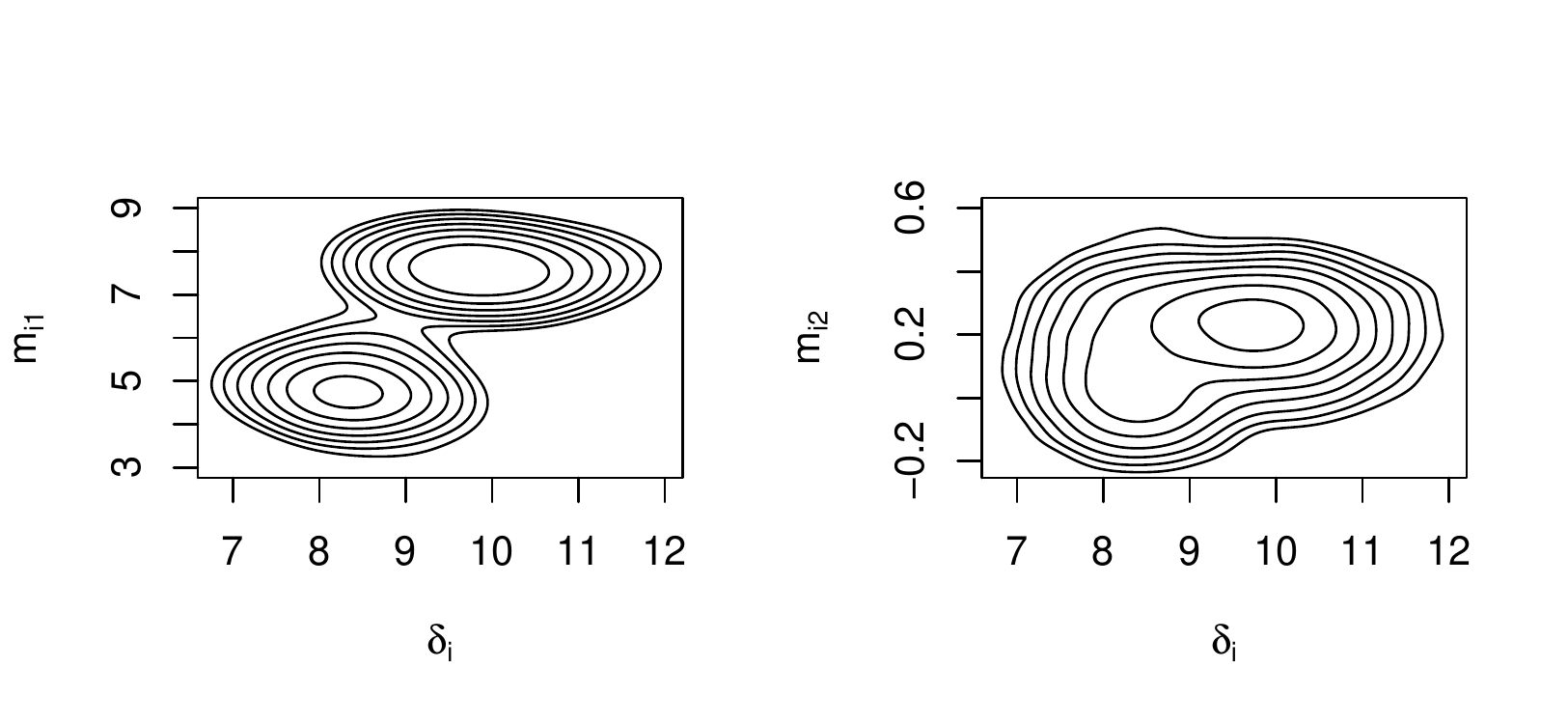}
\caption{
Contour plots of the log of the bivariate posterior predictive densities of $(m_{i1},\, \delta_i)$ (left) and $(m_{i2},\, \delta_i)$ (right) for a hypothetical new patient from the colorectal cancer data.
}
\label{fig:readmission_m_new}
\end{figure}

As discussed in Section~\ref{s:intro}, the DP prior on $(\bm m_i,\,\delta_i)$ described in Section~\ref{sec:prior_spec} allows for clustering of patients based on their recurrent event and survival profiles.
The random effects parameters determine the clustering of patients and capture the dependence between the recurrence and survival processes.
Indeed, the posterior predictive distribution of these parameters for a hypothetical new patient is multimodal as shown in Figure~\ref{fig:readmission_m_new}, indicating the presence of multiple patient subpopulations.
The clustering depends on both gap time trajectories and survival outcomes thanks to the joint distribution on 
 $\bm m_i$ and $\delta_i$. In particular, Figure~\ref{fig:readmission_m_new} reports the bivariate posterior marginals of $(m_{i1}, \delta_{i})$ and $(m_{i2}, \delta_{i})$, which are \replaced{bimodal}{clearly trimodal}.

\begin{figure}
\centering 
\includegraphics[width=0.8\textwidth]{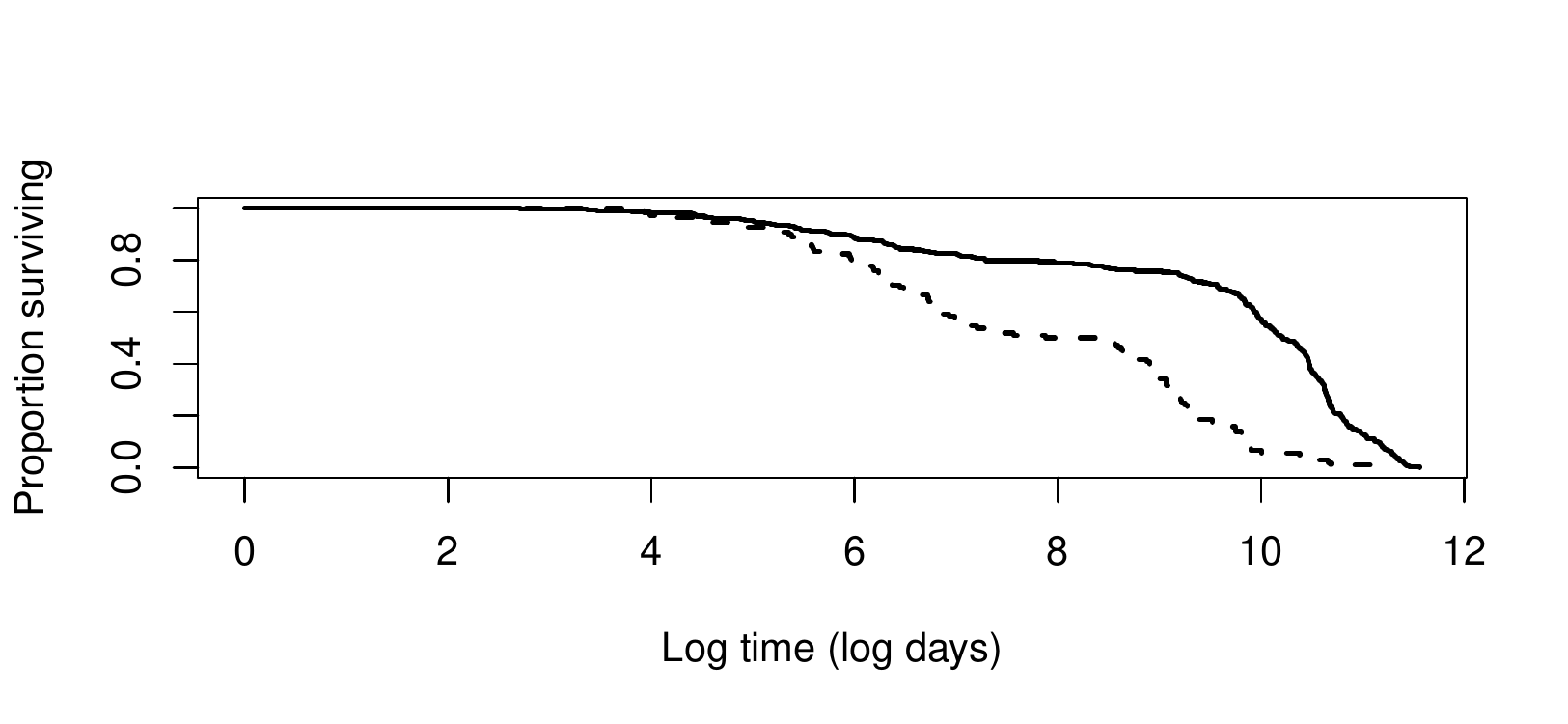}
\caption{
Kaplan-Meier survival estimates for the two largest clusters estimated by 
minimizing the expectation of Binder's\citep{binder1978bayesian}  loss function
under the posterior from our model on the colorectal cancer data.
The solid and dashed lines represent Cluster~1 and 2, respectively.
The curves are based on the posterior means of $\log(S_i)$.
}
\label{fig:readmission_KM}
\end{figure}

Posterior inference on the clustering structure of the patients is of clinical interest as it might guide more targeted therapies.
Our Gibbs sampler provides posterior samples of the cluster allocation.
Here, we report the cluster allocation that minimizes the posterior expectation of Binder's\citep{binder1978bayesian} loss function under equal misclassification costs, which is a common choice in the applied Bayesian non-parametrics literature.\citep{lau2007bayesian}
See Appendix~B of Argiento et al.\cite{argiento2014density} for computational details. Briefly, Binder's loss function measures
the difference for all possible pairs of individuals between the true probability of co-clustering and
the estimated cluster allocation.
In this context, the posterior estimate of the partition of the patients has \added{three} clusters, with \added{99\%} of the patients allocated to two clusters
which are summarized in Table~\ref{tab:readmission_K}.
\begin{table}
\caption{
Summary statistics of the colorectal cancer data and the posterior from our model. The two clusters are the largest from a posterior estimate of the cluster allocation that minimizes the posterior expectation of Binder's\citep{binder1978bayesian} loss function.
The averages and standard deviations of posterior means are taken across patients and recurrent events.
$S_i$ is recorded in days and $Y_{ij}$ in log days.
\label{tab:readmission_K}
}
\centering
\begin{tabular}{lccc}
\hline
& \textbf{Full dataset} & \textbf{Cluster 1} & \textbf{Cluster 2} \\
\hline
Number of patients & 403 & 292 & 108 \\
Proportion censored & 73\% & 81\% & 54\% \\
\hline
Average uncensored $N_i$ & 1.78 (2.92) & 0.79 (0.93) & 2.68 (3.41) \\
Average posterior mean of $N_i$ (SD) & 2.87 (1.95) & 2.55 (1.08) & 3.66 (2.91) \\
Average uncensored $Y_{ij}$ (SD) & 4.35 (1.84) & 5.55 (1.24) & 3.91 (1.72) \\
Average posterior mean of $Y_{ij}$ (SD) & 6.55 (1.39) & 7.06 (0.82) & 5.03 (1.52) \\
Average uncensored $\log(S_i)$ (SD) & 5.78 (1.03) & 5.66 (1.09) & 5.96 (0.93) \\
Average posterior mean of $\log(S_i)$ (SD) & 8.91 (2.14) & 9.39 (2.03) & 7.72 (1.85) \\
\hline
Proportion on chemotherapy & 54\% & 55\% & 50\% \\
Proportion female & 41\% & 42\% & 37\% \\
Proportion stage C & 37\% & 34\% & 45\% \\
Proportion stage D & 19\% & 19\% & 17\% \\
\hline
\end{tabular}
\end{table}
The large\add{st} cluster, Cluster~1,
has longer gap and \added{imputed} survival times than Cluster~2.
Moreover, the Kaplan-Meier curves of each cluster in Figure~\ref{fig:readmission_KM} support the conclusion 
that Cluster~1 includes patients with longer survival times than Cluster~2.
As shown in Table~\ref{tab:readmission_K}, Cluster~1 has a higher censoring rate than Cluster~2, as one might expect at longer survival times.
The lower prevalence of \added{late stage tumours in} Cluster~1 also confirm\added{s} that it includes healthier subjects than Cluster~2.

Figures~\ref{fig:readmission_variance} and \ref{fig:readmission_K} in the supplemental material contain additional posterior results.
\added{Section~\ref{sec:AF} of the supplemental material presents an application to atrial fibrillation data.}

\section{Comparison with other models}
\label{s:comparison}

\subsection{Cox proportional hazards model}

We now compare our results on the colorectal cancer data to those obtained from the Cox proportional hazards model which is one of the most popular semi-parametric models in survival analysis with covariates. In this model, the hazard function for mortality is
${h_i(t\mid \bm{\theta})}= h_0(t)\,\exp({\bm{z}_i^T\bm{\theta}})$
where $h_0(t)$ is the baseline hazard function, $\bm{z}_i$ is a vector of covariates and $\bm{\theta}$ is a vector of regression coefficients.
Here, a larger value of $\bm{z}_i^T\bm{\theta}$ leads to shorter survival times.
This contrasts with \eqref{eq:surv_times} from our model where a larger $\bm{x}_i^T\bm{\gamma}$ is associated with longer survival times.

Our model allows for dependence between the gap and survival times.
Therefore, for a fairer comparison when fitting the Cox proportional hazard model, we include a patient's log mean gap time in the covariate vector $\bm z_i$ in addition to the \added{four} covariates included in $\bm x_i$ described in Section~\ref{sec:set-up}.
\added{As a result, the 199 individuals with no observed gap times are excluded.}
Table~\ref{tab:readmission_PH} in the supplemental material
shows the covariate effects on survival
from the Cox proportional hazards model.
The \added{statistically significant} effects agree with those from our model in Figure~\ref{fig:readmission_CI}.

\subsection{Joint frailty model}

We also compare our model with the joint frailty model by Rondeau et al.\cite{Rondeau2007} as implemented in the R package \texttt{frailtypack}.\citep{Rondeau2012}
The model estimates the hazard functions of rehospitalization and mortality jointly using two patient-specific frailty terms, namely $u_i$ and $v_i$.
The frailty term $u_i$ captures the association between rehospitalization and mortality
while $v_i$ appears solely in the rehospitalization rate.
Specifically, the hazard functions are
$r_{i}(t\mid u_i, v_i, \bm\beta)= u_i\, v_i\, r_0(t)\,\exp({\bm{x}_{i}^T\bm{\beta}})$
for rehospitalization
and
$\lambda_{i}(t|u_i, \bm\gamma)= u_i\,\lambda_0(t)\,\exp({\bm{x}_{i}^T\bm{\gamma}})$
for mortality.
Here, $r_0(t)$ and $\lambda_0(t)$ are baseline hazard functions, and $\bm{x}_{i}$, $\bm\beta$ and $\bm\gamma$
are defined as in Section~\ref{s:model}.
The random effects distributions are specified as follows: $v_i\sim \text{Gamma}(1/\rho,\, 1/\rho)$
and $u_i\sim \text{Gamma}(1/\epsilon,\, 1/\epsilon)$ independently for $i=1,\dots,L$.

The comparison of the joint frailty model results in Table~\ref{tab:readmission_frailty} of the supplemental material with our results in Figure~\ref{fig:readmission_CI}
shows that both models find \added{that late stage tumours are associated with shorter survival. For the other covariates, the comparison is inconclusive with the joint frailty model finding statistically significant effects where our model does not and vice versa.}

Finally, the estimate of $\rho$ is $0.00\added{7}$ with a standard error of $0.00\added{1}$. This suggests heterogeneity between patients that is not explained by the covariates.
The estimate of $\epsilon$ is $0.0\added{73}$ with a standard error of $0.\added{13}$.
This implies that the rate of rehospitalizations is positively associated with mortality.
These results are in line with the posterior clustering results from our model in Table~\ref{tab:readmission_K}
where Cluster~1 is characterized by both the longest gap times and the longest survival times.

\subsection{Bayesian semi-parametric model from Paulon et al.\cite{Paulon2018}}
\label{sec:paulon}

For a more direct comparison, we consider the method proposed by Paulon et al.\cite{Paulon2018}
as it models the gap and survival times jointly using Bayesian non-parametric priors.
Paulon et al.\cite{Paulon2018}
assume that, conditionally on all parameters and random effects, the gap times are independent of both each other and the survival time.
This contrasts with the temporal dependence between gap times in \eqref{eq:gap_times}.
Shared random effects induce dependence between different gap times of the same patient.
Specifically, Paulon et al.\cite{Paulon2018} assume 
$Y_{ij} \sim \mathcal{N}({\bm x}_{i}^T {\bm\beta} + \alpha_{i} ,\, \sigma_i^2)$
independently for $j = 1,\dots,n_i+1$ and $i=1,\dots,L$,
and
\begin{equation} \label{eq:surv_times_Paulon}
	\log (S_i) \sim \mathcal{N}({\bm x}_{i}^T {\bm\gamma} + \psi\,\alpha_{i},\, \eta^2),
\end{equation}
independently for $i=1,\dots,L$,
where $\bm{x}_i$, $\bm{\beta}$ and $\bm{\gamma}$ are defined as in Section~\ref{s:model}, and $\alpha_i$ and $\sigma_i$ are random effects.
Paulon et al.\cite{Paulon2018} do not model the total number of gap times $N_i$ but assume that each patient has a censored $(n_i+1)$th log gap time $Y_{i(n_i+1)}$.
They also assume a priori independence among $\bm{\beta}$, $\bm{\gamma}$, $\psi$, $\eta$ and $(\bm\alpha,\, \bm\sigma^2)$. The random effects $(\alpha_i,\, \sigma_{i}^2) \sim G$ independently for $i=1\dots,L$ where $G\sim\text{DP}(M,\, G_0)$
with $M\sim \text{Uniform}(a_M,\, b_M)$ and
$G_0 = \mathcal{N}(0,\,\alpha_0^2) \times \text{Inv-Gamma}(a_\sigma,\, b_\sigma)$.
The priors on $\bm\beta$, $\bm\gamma$ and $\eta^2$ are set as in Section~\ref{sec:prior_spec}.
Finally, ${\psi} \sim \Nc(0,\, \psi_0^2 )$.

In fitting this model to the colorectal cancer data, we specify the same $\bm x_i$ and the same hyperparameters for the priors on $\bm\beta$, $\bm\gamma$ and $\eta^2$ as in Section~\ref{sec:set-up}.
Furthermore, we set $a_M = 0.3$, $b_M = 5$, $\alpha_0^2 = 100$, $a_\sigma= 2.01$, $b_\sigma =1.01$ and $\psi_0^2 = 100$.
This model yields conclusions that are \added{largely} consistent with those from our model.
In particular, the posterior distributions on the coefficients in Figure~\ref{fig:readmission_CI_paulon} of the supplemental material mimic our results in Figure~\ref{fig:readmission_CI} \added{except for the effect of tumour stage on recurrence}.
Also, the posterior on $\psi$ in \eqref{eq:surv_times_Paulon} concentrates between \added{1.3 and 1.7} per Figure~\ref{fig:readmission_psi} of the supplemental material.
This parameter captures the strength of the relationship between gap and survival times.
Thus, the time between hospitalizations and survival have a positive association.
This is consistent with the clustering results obtained from our model in Table~\ref{tab:readmission_K}.
Lastly, the posterior on the number of clusters for this model and our model vary slightly, with a mode of \added{three} clusters for our model in Figure~\ref{fig:readmission_K} of the supplemental material while Figure~\ref{fig:readmission_K_paulon} of the supplemental material has the mode at \added{five} for the model from Paulon et al.\cite{Paulon2018}
This is not surprising as our model introduces more structure: a temporal model for the gap time, as well as process-specific frailty terms which are jointly modelled non-parametrically.  Moreover, the number of recurrent events is a random quantity and object of inference, which also informs the dependence between the two processes, in addition to the truncation $T_{iN_i} \leq S_i$ of \eqref{eq:gap_times}.
In contrast, Paulon et al.\cite{Paulon2018} can capture such dependence using only $\psi$ in \eqref{eq:surv_times_Paulon}, and the gap times are conditionally independent given $\psi$ and the remaining parameters, with the total number of gap times per individual assumed arbitrarily large.

\section{Discussion}
\label{s:discuss}

We introduce a joint model on recurrence and survival that explicitly treats the number of recurrent events $N_i$ before the terminal event as a random variable and object of inference as $N_i$ is often of interest in applications.
Additionally, the explicit modelling of $N_i$ as well as the specification of a joint distribution for the random effects of the recurrence and survival processes
induces dependence between these processes.
Moreover, temporal dependence among recurrent events is introduced through a first-order autoregressive process on the gap times.
Extension to a more complex temporal structure is in principle straightforward.
The model allows for estimating covariates effects on the recurrence and survival processes by introducing appropriate regression terms.
Our model can readily accommodate a different prior on the number of recurrent events than the \replaced{negative binomial}{Poisson} distribution. For instance, \replaced{an earlier version of this work\cite{vandenBoom2020} used a Poisson distribution}{a negative-binomial can be used to allow for overdispersion}.
The use of a non-parametric prior as random effects distribution allows for extra flexibility, patient heterogeneity and data-driven clustering of the patients.

\added{We use log-normal kernels for the non-parametric survival and gap time distributions.
Use of different kernels is computationally impracticable
as then
the evaluation of the normalization constant of
${p(\bm Y_i, S_i\mid \text{\textemdash} )}$ becomes considerably more expensive.
This evaluation is already the computational bottleneck of our method
as it happens frequently as part of slice sampling at each iteration of the Gibbs sampler.
The reasons why using different kernels is problematic are twofold:
Firstly,
the log-normal kernel for the gap times enables the Fenton-Wilkinson\citep{Fenton1960} approximation for the distribution of $T_{iN_i}$.
Other kernels for the gap times might not enable such computationally efficient evaluations.
For instance, Proposition~1 of El Bouanani and Ben-Azza\cite{ElBouanani2015} gives rise to an approximation when using a Weibull kernel which is computationally more involved than the Fenton-Wilkinson method.
Secondly,
the current combination of the Fenton-Wilkinson method
and a log-normal survival kernel
reduce the normalization constant of
${p(\bm Y_i, S_i\mid \text{\textemdash} )}$
to a tail probability of a univariate Gaussian
as detailed in Section~\ref{sec:constant} of the supplemental material.
A different survival kernel, say a Weibull kernel,
would require computation which is an order of magnitude more expensive, such as numerical integration to evaluate the normalization constant,
or further approximation.
}

The simulation study shows the effectiveness of posterior inference on the number of recurrent events $N_i$.
Comparison with the Cox proportional hazards model, the joint frailty model\citep{Rondeau2007} and the Bayesian semi-parametric model from Paulon et al.\cite{Paulon2018} yields consistent results,
with few
exceptions in the estimation of covariate effects.
This discrepancy might be the result of the fact that these models have fewer parameters and assume a single patient population
while our model detects multiple subpopulations.
\add{Moreover, in our model specification a further level of dependence is introduced through the  distribution of the number of gap times. More flexibility, if required by the application, could be achieved by including also the hyper-parameters of the distribution of the number of gap-times in the nonparametric component of the model.}

\begin{funding}
\added{This research is partially supported by the Singapore Ministry of Health’s National Medical Research Council under its Open Fund - Young Individual Research Grant (OFYIRG19nov-0010).}
\end{funding}

\begin{dci}
The Authors declare that there is no conflict of interest.
\end{dci}

\begin{sm}
The supplemental material referenced in the text is available online.
The code that generated the results in this paper is available at \url{https://github.com/willemvandenboom/condi-recur}.
\end{sm}

\bibliographystyle{SageV}
\bibliography{condi-recur}

\begin{thebibliography}{10}
\providecommand{\url}[1]{\texttt{#1}}
\providecommand{\urlprefix}{URL }
\expandafter\ifx\csname urlstyle\endcsname\relax
  \providecommand{\doi}[1]{DOI:\discretionary{}{}{}#1}\else
  \providecommand{\doi}{DOI:\discretionary{}{}{}\begingroup
  \urlstyle{rm}\Url}\fi
\providecommand{\eprint}[2][]{\url{#2}}

\bibitem{Jencks2009}
Jencks SF, Williams MV and Coleman EA.
\newblock Rehospitalizations among patients in the medicare fee-for-service
  program.
\newblock \emph{New England Journal of Medicine} 2009; 360(14): 1418--1428.
\newblock \doi{10.1056/nejmsa0803563}.

\bibitem{McIlvennan2015}
McIlvennan CK, Eapen ZJ and Allen LA.
\newblock Hospital readmissions reduction program.
\newblock \emph{Circulation} 2015; 131(20): 1796--1803.
\newblock \doi{10.1161/circulationaha.114.010270}.

\bibitem{Zuckerman2016}
Zuckerman RB, Sheingold SH, Orav EJ et~al.
\newblock Readmissions, observation, and the hospital readmissions reduction
  program.
\newblock \emph{New England Journal of Medicine} 2016; 374(16): 1543--1551.
\newblock \doi{10.1056/nejmsa1513024}.

\bibitem{Cook2007}
Cook RJ and Lawless JF.
\newblock \emph{The statistical analysis of recurrent events}.
\newblock Springer, New York, 2007.
\newblock \doi{10.1007/978-0-387-69810-6}.

\bibitem{Liu2004}
Liu L, Wolfe RA and Huang X.
\newblock Shared frailty models for recurrent events and a terminal event.
\newblock \emph{Biometrics} 2004; 60(3): 747--756.
\newblock \doi{10.1111/j.0006-341x.2004.00225.x}.

\bibitem{Rondeau2007}
Rondeau V, Mathoulin-Pelissier S, Jacqmin-Gadda H et~al.
\newblock Joint frailty models for recurring events and death using maximum
  penalized likelihood estimation: {A}pplication on cancer events.
\newblock \emph{Biostatistics} 2007; 8(4): 708--721.
\newblock \doi{10.1093/biostatistics/kxl043}.

\bibitem{Ye2007}
Ye Y, Kalbfleisch JD and Schaubel DE.
\newblock Semiparametric analysis of correlated recurrent and terminal events.
\newblock \emph{Biometrics} 2007; 63(1): 78--87.
\newblock \doi{10.1111/j.1541-0420.2006.00677.x}.

\bibitem{Huang2010}
Huang CY, Qin J and Wang MC.
\newblock Semiparametric analysis for recurrent event data with time-dependent
  covariates and informative censoring.
\newblock \emph{Biometrics} 2010; 66(1): 39--49.
\newblock \doi{10.1111/j.1541-0420.2009.01266.x}.

\bibitem{Sinha2008}
Sinha D, Maiti T, Ibrahim JG et~al.
\newblock Current methods for recurrent events data with dependent termination.
\newblock \emph{Journal of the American Statistical Association} 2008;
  103(482): 866--878.
\newblock \doi{10.1198/016214508000000201}.

\bibitem{Ouyang2013}
Ouyang B, Sinha D, Slate EH et~al.
\newblock Bayesian analysis of recurrent event with dependent termination: An
  application to a heart transplant study.
\newblock \emph{Statistics in Medicine} 2013; 32(15): 2629--2642.
\newblock \doi{10.1002/sim.5717}.

\bibitem{Olesen2006}
Olesen AV and Parner ET.
\newblock Correcting for selection using frailty models.
\newblock \emph{Statistics in Medicine} 2006; 25(10): 1672--1684.
\newblock \doi{10.1002/sim.2298}.

\bibitem{Huang2007}
Huang X and Liu L.
\newblock A joint frailty model for survival and gap times between recurrent
  events.
\newblock \emph{Biometrics} 2007; 63(2): 389--397.
\newblock \doi{10.1111/j.1541-0420.2006.00719.x}.

\bibitem{Yu2011}
Yu Z and Liu L.
\newblock A joint model of recurrent events and a terminal event with a
  nonparametric covariate function.
\newblock \emph{Statistics in Medicine} 2011; 30(22): 2683--2695.
\newblock \doi{10.1002/sim.4297}.

\bibitem{Bao2012}
Bao Y, Dai H, Wang T et~al.
\newblock A joint modelling approach for clustered recurrent events and death
  events.
\newblock \emph{Journal of Applied Statistics} 2012; 40(1): 123--140.
\newblock \doi{10.1080/02664763.2012.735225}.

\bibitem{Liu2015}
Liu L, Huang X, Yaroshinsky A et~al.
\newblock Joint frailty models for zero-inflated recurrent events in the
  presence of a terminal event.
\newblock \emph{Biometrics} 2015; 72(1): 204--214.
\newblock \doi{10.1111/biom.12376}.

\bibitem{Yu2013}
Yu Z, Liu L, Bravata DM et~al.
\newblock Joint model of recurrent events and a terminal event with
  time-varying coefficients.
\newblock \emph{Biometrical Journal} 2013; 56(2): 183--197.
\newblock \doi{10.1002/bimj.201200160}.

\bibitem{Li2018}
Li Z, Chinchilli VM and Wang M.
\newblock A {B}ayesian joint model of recurrent events and a terminal event.
\newblock \emph{Biometrical Journal} 2019; 61(1): 187--202.
\newblock \doi{10.1002/bimj.201700326}.

\bibitem{Li2019}
Li Z, Chinchilli VM and Wang M.
\newblock A time-varying {B}ayesian joint hierarchical copula model for
  analysing recurrent events and a terminal event: an application to the
  cardiovascular health study.
\newblock \emph{Journal of the Royal Statistical Society: Series C (Applied
  Statistics)} 2019; 69(1): 151--166.
\newblock \doi{10.1111/rssc.12382}.

\bibitem{Paulon2018}
Paulon G, {De Iorio} M, Guglielmi A et~al.
\newblock Joint modeling of recurrent events and survival: A {B}ayesian
  non-parametric approach.
\newblock \emph{Biostatistics} 2018; \doi{10.1093/biostatistics/kxy026}.
\newblock Kxy026.

\bibitem{Tallarita2020}
Tallarita M, {De Iorio} M, Guglielmi A et~al.
\newblock Bayesian autoregressive frailty models for inference in recurrent
  events.
\newblock \emph{The International Journal of Biostatistics} 2020; 16(1): 1--18.
\newblock \doi{10.1515/ijb-2018-0088}.

\bibitem{Kansagara2011}
Kansagara D, Englander H, Salanitro A et~al.
\newblock Risk prediction models for hospital readmission: A systematic review.
\newblock \emph{{JAMA}} 2011; 306(15): 1688.
\newblock \doi{10.1001/jama.2011.1515}.

\bibitem{Futoma2015}
Futoma J, Morris J and Lucas J.
\newblock A comparison of models for predicting early hospital readmissions.
\newblock \emph{Journal of Biomedical Informatics} 2015; 56: 229--238.
\newblock \doi{10.1016/j.jbi.2015.05.016}.

\bibitem{Mahmoudi2020}
Mahmoudi E, Kamdar N, Kim N et~al.
\newblock Use of electronic medical records in development and validation of
  risk prediction models of hospital readmission: systematic review.
\newblock \emph{{BMJ}} 2020; 369: m958.
\newblock \doi{10.1136/bmj.m958}.

\bibitem{Green1995}
Green PJ.
\newblock Reversible jump {M}arkov chain {M}onte {C}arlo computation and
  {B}ayesian model determination.
\newblock \emph{Biometrika} 1995; 82(4): 711--732.
\newblock \doi{10.1093/biomet/82.4.711}.

\bibitem{Neal2003}
Neal RM.
\newblock Slice sampling.
\newblock \emph{The Annals of Statistics} 2003; 31(3): 705--767.
\newblock \doi{10.1214/aos/1056562461}.

\bibitem{Ferguson1973}
Ferguson TS.
\newblock A {B}ayesian analysis of some nonparametric problems.
\newblock \emph{The Annals of Statistics} 1973; 1(2): 209--230.

\bibitem{Sethuraman1994}
Sethuraman J.
\newblock A constructive definition of {D}irichlet priors.
\newblock \emph{Statistica Sinica} 1994; 4(2): 639--650.

\bibitem{Aalen1991}
Aalen OO and Husebye E.
\newblock Statistical analysis of repeated events forming renewal processes.
\newblock \emph{Statistics in Medicine} 1991; 10(8): 1227--1240.
\newblock \doi{10.1002/sim.4780100806}.

\bibitem{Lo1984}
Lo AY.
\newblock On a class of {B}ayesian nonparametric estimates: I. density
  estimates.
\newblock \emph{The Annals of Statistics} 1984; 12(1).
\newblock \doi{10.1214/aos/1176346412}.

\bibitem{DeIorio2009}
{De Iorio} M, Johnson WO, M\"{u}ller P et~al.
\newblock Bayesian nonparametric nonproportional hazards survival modeling.
\newblock \emph{Biometrics} 2009; 65(3): 762--771.
\newblock \doi{10.1111/j.1541-0420.2008.01166.x}.

\bibitem{Fenton1960}
Fenton L.
\newblock The sum of log-normal probability distributions in scatter
  transmission systems.
\newblock \emph{{IEEE} Transactions on Communications} 1960; 8(1): 57--67.
\newblock \doi{10.1109/tcom.1960.1097606}.

\bibitem{Neal2000}
Neal RM.
\newblock {M}arkov chain sampling methods for {D}irichlet process mixture
  models.
\newblock \emph{Journal of Computational and Graphical Statistics} 2000; 9(2):
  249--265.
\newblock \doi{10.1080/10618600.2000.10474879}.

\bibitem{Gonzalez2005}
Gonz{\'a}lez JR, Fernandez E, Moreno V et~al.
\newblock Sex differences in hospital readmission among colorectal cancer
  patients.
\newblock \emph{Journal of Epidemiology {\&} Community Health} 2005; 59(6):
  506--511.
\newblock \doi{10.1136/jech.2004.028902}.

\bibitem{Rondeau2012}
Rondeau V, Mazroui Y and Gonzalez JR.
\newblock {frailtypack}: An {R} package for the analysis of correlated survival
  data with frailty models using penalized likelihood estimation or
  parametrical estimation.
\newblock \emph{Journal of Statistical Software} 2012; 47(4): 1--28.
\newblock \doi{10.18637/jss.v047.i04}.

\bibitem{binder1978bayesian}
Binder DA.
\newblock Bayesian cluster analysis.
\newblock \emph{Biometrika} 1978; 65(1): 31--38.

\bibitem{lau2007bayesian}
Lau JW and Green PJ.
\newblock Bayesian model-based clustering procedures.
\newblock \emph{Journal of Computational and Graphical Statistics} 2007; 16(3):
  526--558.

\bibitem{argiento2014density}
Argiento R, Cremaschi A and Guglielmi A.
\newblock A ``density-based'' algorithm for cluster analysis using species
  sampling {G}aussian mixture models.
\newblock \emph{Journal of Computational and Graphical Statistics} 2014; 23(4):
  1126--1142.

\bibitem{vandenBoom2020}
van~den Boom W, Tallarita M and {De Iorio} M.
\newblock {B}ayesian joint modelling of recurrence and survival: a conditional
  approach, 2020.
\newblock {arXiv:2005.06819v1}.

\bibitem{ElBouanani2015}
Bouanani FE and Ben-Azza H.
\newblock Efficient performance evaluation for {EGC}, {MRC} and {SC} receivers
  over {W}eibull multipath fading channel.
\newblock In \emph{Lecture Notes of the Institute for Computer Sciences, Social
  Informatics and Telecommunications Engineering}. Springer International
  Publishing, 2015.
\newblock pp. 346--357.
\newblock \doi{10.1007/978-3-319-24540-9_28}.

\end{thebibliography}

\includepdf[scale=1,pages=-]{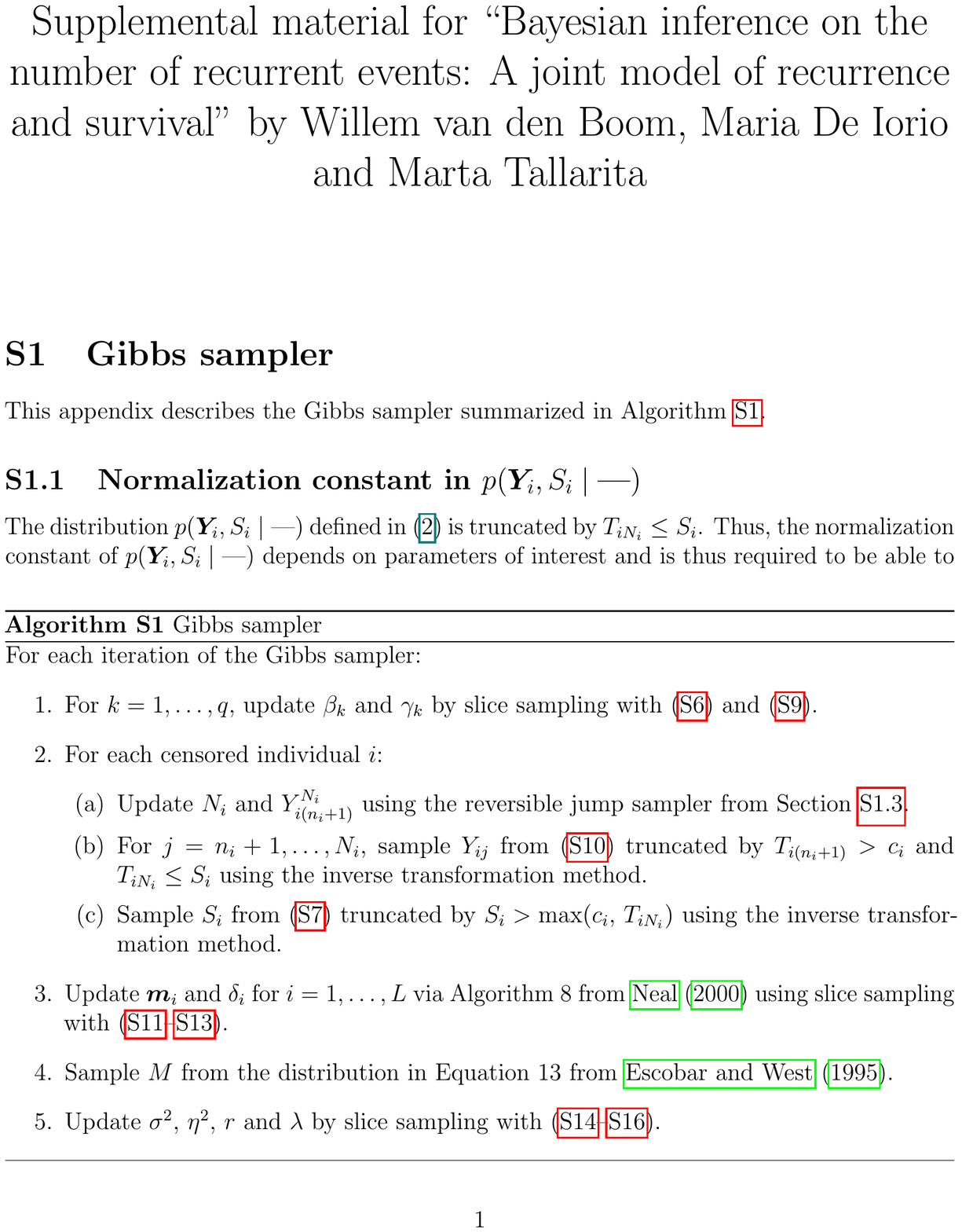}

\end{document}